\documentclass[prb]{revtex4-2}

\usepackage{amsmath}
\usepackage{amsfonts}
\usepackage{graphicx}
\usepackage{hyperref}

\newcommand\dd{\mathrm{d}}   

\renewcommand{\vec}{\mathbf} 

\newcommand{\uvec}[1]{\vec{\hat #1}}  

\newcommand{\uvx}{\uvec x}
\newcommand{\uvy}{\uvec y}
\newcommand{\uvz}{\uvec z}

\newcommand{\mcs}{{\mathrm{mcs}}}
\newcommand{\tens}{{\mathrm{tensile}}}
\newcommand{\shear}{{\mathrm{shear}}}
\newcommand{\bend}{{\mathrm{bending}}}
\newcommand{\tors}{{\mathrm{torsion}}}
\newcommand{\free}{{\mathrm{free}}}
\newcommand{\fixed}{{\mathrm{fixed}}}

\begin{document}

\title{The nonlinearity of helical springs: An energy-based approach}

\author{Saša Ilijić}
\author{Ana Babić} \email{ana.babic@fer.hr} 
\author{Dora Ivrlač}
\affiliation{University of Zagreb 
       Faculty of Electrical Engineering and Computing,
       Department of Applied Physics, Unska 3, 10\,000 Zagreb, Croatia}

\author{Andrew DeBenedictis}
\affiliation{Simon Fraser University Faculty of Science, and
       The Pacific Institute for the Mathematical Sciences,
       Burnaby, British Columbia, V5A 1S6, Canada}

\date{\today}

\begin{abstract}
A general expression for the elastic potential energy
of a helical spring is derived
using the basic concepts of elasticity theory and geometry.
Both the translational and the rotational displacement
of the spring's moving ends are considered.
The resulting expression is employed to derive the general relation
between force and spring height,
and corrections to Hooke's law are discussed.
The results can be used to estimate the magnitude
of non-linear effects in realistic situations,
to predict how springs unwind under load,
and to determine the spring constant
when spring ends are allowed to rotate freely.
This energy-based approach is suitable for inclusion
in advanced mechanics courses.
The results are validated via measurements on steel springs
using a simple experimental setup
appropriate for use in physics laboratories at all levels. 
\end{abstract}

\maketitle


\section{Introduction and preliminaries}

Helical springs are common in both mechanical systems
and physics laboratories.
Expressions relating the spring stiffness
to the geometry and the physical properties of the spring material
have been derived in textbooks and physics education papers.
%
%
Early treatments addressed
only small spring deflections,%
   \cite{*[{}] [{, pp.~415--417.}] lovebook}%
   \cite{*[{}] [{, p.~71, pp.~77--81.}] timoshenkobook}
while more recent pedagogical works
extend the analysis beyond the linear regime
and usually do not discuss
the unwinding of the spring.\cite{mohazzabi89,ivchenko20}
Some engineering approaches
focus on closely coiled springs\cite{kato21}
or provide series expansions for spring stiffness
around the relaxed state.\cite{jiang}
A thorough treatment of spring mechanics
may be found in the book by Wahl.%
\cite{*[{}] [{, pp.~50--68.}] wahlbook}
Most of these treatments analyze
local forces and torques within the spring material,
and only some of them derive the expressions
for the energy of the spring.%
\cite{*[{}] [{, pp.~308--314.}] sommerfeldbook}%
\cite{huang24}
However, modern classical mechanics courses
often prefer an energy-based approach.

Here, we derive the general expression
for the elastic potential energy of a helical spring
using exclusively the concepts of elastic potential energy
and the spring's geometry,
without starting from forces and torques.
The potential energy is then employed
to find the general relation between spring force and height,
as well as to predict
how a spring with freely rotating ends unwinds under load.
We see several advantages this work may provide:
(i) It naturally leads to a discussion of the difference
between the maximally compressed state and the relaxed state of a spring,
giving a method to obtain information on the relaxed state
even if it is physically inaccessible.
(ii) It provides a complete treatment of the (un)winding of the spring,
valid for any initial or subsequent pitch angle.
In particular, it is applicable to closely or loosely coiled springs,
both in compression and extension.
(iii) The exact expressions for potential energy and spring force
enable series expansions around any desired working point,
such as vertical oscillations of a mass on a spring
or of a heavily loaded, loosely coiled automotive suspension spring.

The work was motivated by the design
of a student laboratory experiment
to investigate resonance in coupled oscillators.
Due to the weak damping in such systems,
the resonance is narrow, and we needed the precise evolution 
of spring stiffness with load.
This work could however be useful in other contexts.
In modern physics laboratories,
where electromechanical sensors, actuators,
and digital signal processing enable high-precision measurements,
deviation from Hooke's law can be measured for instance in experiments
studying transients in forced oscillations,
oscillations in systems with more than one degree of freedom,
or a spring's larger elongations. \\

\begin{figure}
\begin{center}
\includegraphics[angle=-90]{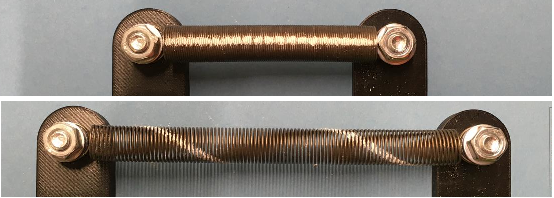}
\end{center}
\caption{\label{fig:chalk}
The upper photo shows the spring in its maximally compressed state
with a chalk line marking parts of its outer surface.
In the lower photo the same spring has been stretched and twisted by $4\pi$.
Note that the chalk line remains on the outer surface of the spring
indicating the validity of our assumption 5.}
\end{figure}

A spring is said to be linear if its restoring force
can be expressed as $F(x) = - k ( x - x_0 )$,
where $x$ is the coordinate of spring's moving end,
$x_0$ is a constant, and $k$ is the spring's coefficient of elasticity.
Equivalently, a spring is said to be linear
if the potential energy associated with the spring force 
is of the form $U(x) = (k/2) (x-x_0)^2 $,
or if the coefficient of elasticity
\begin{equation}
k = - \lim_{\Delta x \to 0} \frac{\Delta F}{\Delta x}
  = - \frac{\dd}{\dd x} F(x) = \frac{\dd^2}{\dd x^2} U(x)
\end{equation}
does not depend on $x$.
It is then commonly known as the spring constant.
However, real-life springs are non-linear.
To build a realistic model that describes the spring behavior
beyond the linear regime, we make the following assumptions:
\begin{enumerate}

\item The deformation of the wire material that constitutes the spring
      remains within the regime of linear elasticity at all spring loads.
      The nonlinearity of interest here arises
      from the geometry of the spring.\footnote{
      It is difficult to find a proper spring steel stress-strain curve.
      For similar steels, the measured curves indicate
      that the non-linear elastic regime
      and the plastic deformation regime are close.
      One could argue that if the steel is in the non-linear regime,
      it is close to being permanently deformed,
      so that we believe this assumption is realistic,
      provided the spring is not deformed during use.}

\item As the spring extends or contracts under varying load,
      the spring wire undergoes torsion and bending,
      while the length of the wire is assumed to remain unchanged.
      The cross-section of the wire is assumed
      to remain circular at all spring loads.\cite{timoshenkobook}

\item All calculations are carried out to leading order
      of the power expansions in the parameter $d/R_0$,
      where $d$ is the wire diameter,
      and $R_0$ is the wire curvature radius
      (or approximately half of the spring diameter).
      This is known as the ``thin wire approximation''\cite{ivchenko20}
      and is briefly discussed in Appendix \ref{app:bending}.

\item If only one end of the spring is fixed,
      then the other end can move along the spring's longitudinal axis,
      but it can also rotate about that axis,
      thus winding up or unwinding the spring.
      We will consider both of these degrees of freedom for the spring.

\item The parts of the spring wire that are on the outer (inner) radius
      remain on the outer (inner) radius at all spring loads.
      The validity of this assumption is illustrated
      in Fig.~\ref{fig:chalk}.
      This observation will enable us to attach
      a comoving coordinate system all along the spring wire material,
      and to compute the amount of torsion
      the spring wire undergoes as the spring deforms.

\end{enumerate}

\begin{figure}
\begin{center}
\includegraphics{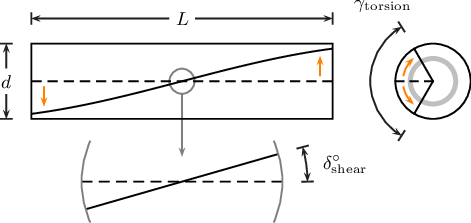}
\end{center}
\caption{\label{fig:torsion}
Wire element of length $L$ and diameter $d$ subject to torsion.
Left: Side view of the wire.  Right: Cross-section of the wire.
The points initially located along the dashed line in the relaxed wire
are displaced into a helix (solid curve) when the wire is under torsion.
The opposite directions of rotation
of the two wire ends are indicated by arrows.
The wire torsion angle $\gamma_\tors$
and the wire surface shear angle $\delta^{\circ}_\shear$
are related by $\gamma_\tors = (2L/d) \delta^\circ_\shear$.
The shaded region of the cross-section indicates
the surface element used in the integration of the potential energy
due to torsion, see Appendix~\ref{app:torsion}.}
\end{figure}

From the theory of elasticity,
we use the expression for the volume density
of elastic potential energy $u_\shear$ under shear stress,
\begin{equation} \label{eq:u.shear}
u_\shear = \frac12 G \delta_\shear^2,
\end{equation}
where $\delta_\shear$ is the shear angle,
$G$ is the shear modulus of the material,
and the quadratic dependence of $u_\shear$ on $\delta_\shear$
implies linear elasticity of the material.%
\cite{*[{}] [{, pp.~140--143.}] masebook}%
\cite{sommerfeldbook}
This allows one to compute
the elastic potential energy $U_\tors$
of a wire of length $L$ and diameter $d$ whose ends are
rotated about the wire center line
by the angle $\gamma_\tors$, see Fig.~\ref{fig:torsion}.
The elastic energy due to torsion is
\begin{equation} \label{eq:ut.gen}
U_\tors = G \frac{(d/2)^4 \pi}{4L} \gamma_\tors^2 ,
\end{equation}
where it is assumed that in the relaxed state of the wire
$\gamma_\tors = 0$.
The derivation of Eq.~(\ref{eq:ut.gen}) is given in Appendix \ref{app:torsion}.

\begin{figure}
\begin{center}
\includegraphics{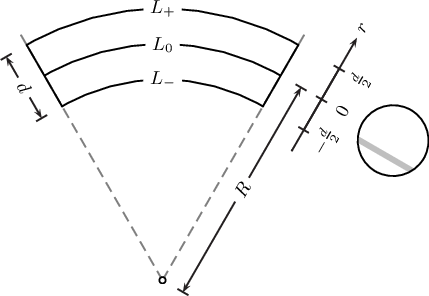}
\end{center}
\caption{\label{fig:bending}
Element of a curved wire with diameter $d$,
center line length $L_0$, and center line curvature radius $R$:
The center line length $L_0$ is assumed to
be invariant upon bending, that is under changes in $R$,
while the lengths of the innermost and the outermost arcs
$L_\pm$ change with $R$.
The circle to the right is the cross-section of the wire,
the shaded region indicates the surface element used in the integration
of potential energy due to bending, see Appendix (\ref{app:bending}).}
\end{figure}

The second result we draw from elasticity
is the expression for the volume density
of the elastic potential energy $u_\tens$
in a material subject to tensile or compressive stress,
valid in the linear elastic regime,
\begin{equation} \label{eq:u.tens}
u_\tens = \frac12 E \delta_L^2,
\end{equation}
where $\delta_L$ is the relative length deformation
of the material in the direction of the applied stress
and $E$ is Young's modulus of the material.\cite{masebook,timoshenkobook}
This allows one to derive the expression
for the elastic potential energy $U_\bend$
of a wire of length $L$ and diameter $d$
which is bent so that its curvature radius is $R$, see Fig.~\ref{fig:bending}.
If in the relaxed state its curvature radius is $R_0$,
then, within the thin wire approximation,
the wire's elastic potential energy is
\begin{equation} \label{eq:ub.gen}
U_\bend
  = E \frac{(d/2)^4 \pi L}{8 R_0^2} \left( \frac{R-R_0}{R} \right)^2.
\end{equation} 
The derivation of Eq.~(\ref{eq:ub.gen})
is outlined in Appendix \ref{app:bending}.

In the following sections we first consider the geometry of the spring
in order to derive the expressions for
torsion and for the spring's curvature radius
that will be used in Eqs~\ref{eq:ut.gen}) and (\ref{eq:ub.gen}).
In Sections~\ref{sec:uspring} and \ref{sec:force}
we derive the complete expression
for the elastic potential energy of the spring
and explore some of its predictions.
Experimental verification of some of our results
is given in Sec.~\ref{sec:experiment}.


\section{The geometry of the spring \label{sec:geometry}}

\begin{figure}
\begin{center}
\includegraphics{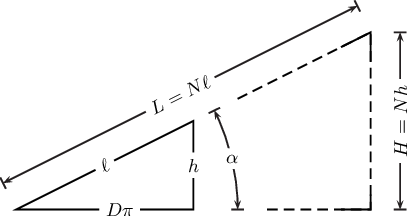}
\end{center}
\caption{\label{fig:triangles}
Diagram clarifying the notations used in this paper:
The hypotenuse of the small triangle represents one coil of the spring,
$\ell^2 = (D\pi)^2 + h^2$.
The hypotenuse of the large triangle (not to scale, dashed)
represents the whole spring.} 
\end{figure}

\begin{figure}
\begin{center}
\includegraphics{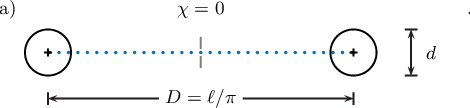}\\[2em]
\includegraphics{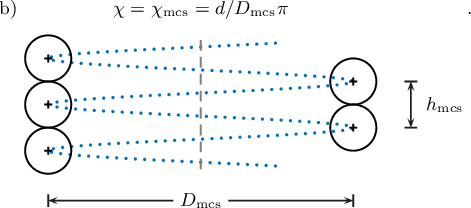}\\[2em]
\includegraphics{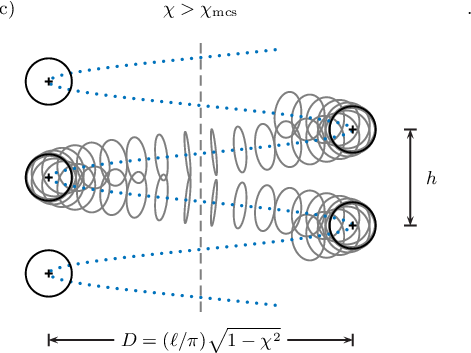}
\end{center}
\caption{ \label{fig:xsections} (Color online)
Longitudinal cross-section of a helical spring
with wire of diameter $d$ and length of wire per turn $\ell$.
Black circles (which are actually slightly elliptical in b) and c))
are the intersections of the wire with the plane of the drawing.
Blue dotted curves are the projections of the wire center line
onto the plane of the drawing.
Gray dashed lines are the longitudinal axes of the spring.
a) Spring in the physically inaccessible state
   with overlapping turns of wire, $\chi=0$.
b) Spring in the maximally compressed state ($\mcs$).
c) Spring in a realistic state with $\chi=1/10$.
   Gray ellipses are the projections of the
   wire cross-sections onto the plane of the drawing.
}
\end{figure}

As previously mentioned,
we will restrict ourselves to the case of a spring
whose wire has a diameter $d$ and and total length $L$,
both of which are assumed not to change with load.
The number $N$ of turns in the spring
is considered to be a continuous variable
since we allow the spring to wind or unwind.
The length $\ell$ of wire per turn, $\ell = L/N$,
is therefore variable as well.
We will be referring to spring's length
as ``spring's height'' and we denote it with $H$.
The height $h=H/N$ of one coil is known as the coil pitch.
As we will not be considering the effect of spring's own weight,
the coil pitch is assumed to be uniform along the spring.
We will also extensively use the ratio
\begin{equation} \label{eq:chidef}
\chi = \frac{h}{\ell} = \frac{H}{L} = \sin\alpha,
\end{equation}
which we call the pitch parameter, where $\alpha$ is the usual pitch angle.
A graphical overview of the notation is given in Fig.~\ref{fig:triangles}.

The pitch parameter $\chi = 0$ corresponds to completely overlapping coils,
a state which is useful to consider
even though it is physically inaccessible.
It is illustrated in Fig.~\ref{fig:xsections}-a,
in which the spring has the shape of a torus.
The maximal value of the pitch parameter, $\chi = 1$,
corresponds to a completely straight wire.

The spring at maximum compression,
in which the turns of wire touch one another,
but do not overlap, is shown in Fig.~\ref{fig:xsections}-b.
Simple geometric considerations reveal
that the pitch parameter in this state is
\begin{equation}
\chi_\mcs = \frac{h_\mcs}{\ell_\mcs}
= \frac{H_\mcs}{L} = \frac{d}{D_\mcs \pi}.
\end{equation}

The relaxed state of the spring, denoted by $\chi_0$ and $N_0$,
is the one in which there is no internal stress in the wire's material.
Here we must distinguish two classes of springs:
If $\chi_0 > \chi_\mcs$,
the spring can both extend and compress relative to its relaxed state,
and, in the absence of external force, it rests in its relaxed state.
Such springs are used, for example, in car suspensions.
The other class of springs has $\chi_0 < \chi_\mcs$,
meaning that they cannot reach their relaxed state.
Without external forces, they remain in their maximally compressed states.
For example, such springs are commonly used in classroom demonstrations.

Assuming that one end of the spring is fixed,
we consider two degrees of freedom for the moving end.
The first one is the motion along the longitudinal axis of the spring,
i.e.\ the compression or elongation of the spring,
which can be parameterized by the pitch parameter $\chi$
defined in Eq.~(\ref{eq:chidef}).
The second is the rotation of the moving end about the spring axis,
i.e.\ the winding or unwinding of the spring,
which can be parameterized by
the number of turns $N$ of the wire in the spring.
If the moving end of the spring rotates by $\Delta\phi$,
then $\Delta N = \Delta \phi / 2\pi$
(recall that $N$ is a continuous variable).

We begin modeling the spring by introducing
the helix that represents the center line of the spring wires
(blue solid curve in Figs \ref{fig:xsections} and \ref{fig:vectors}).
Taking the $z$-axis along the spring's longitudinal axis,
this helix is parameterized as
\begin{equation} \label{eq:wirecl}
\vec r(s) = \frac{\ell}{2\pi} \sqrt{1-\chi^2}
              \left( \cos\frac{2\pi s}\ell\;\uvx
                   + \sin\frac{2\pi s}\ell\;\uvy \right) + s \chi \; \uvz,
\end{equation}
where the parameter $s$ is the wire's length from one of its ends,
$0\le s\le L$, see Fig.~\ref{fig:vectors}.
In this parameterization of the helix,
a fixed value of $s$ corresponds to a specific material particle
on the wire's center line for all spring loads.
The parameter $s$ can therefore be understood
as the attached-to-material or comoving
coordinate running along the length of the wire.

We introduce local orthonormal basis vectors
at points on the wire's center line, see Fig.~\ref{fig:vectors}.
As the first basis vector we choose the vector $\uvec w$
that is tangential to the wire center line.
This vector is obtained by differentiating $\vec r$
with respect to the curve length $s$ giving the unit vector
\begin{equation} \label{eq:w}
\uvec w(s) = \frac{\dd \vec r}{\dd s}
           = \sqrt{1 - \chi^2}
             \left( - \sin\frac{2\pi s}{\ell} \; \uvx
                    + \cos\frac{2\pi s}{\ell} \; \uvy \right)
           + \chi \; \uvz .
\end{equation}
The remaining two basis vectors are orthogonal to $\uvec w$
and lie in the plane of the wire's
cross-section as shown in Fig.~\ref{fig:vectors}.
As the first of these vectors we choose
\begin{equation} \label{eq:v}
\uvec v(s) = \cos\frac{2\pi s}{\ell} \; \uvx + \sin\frac{2\pi s}{\ell} \; \uvy,
\end{equation}
which points outward relative to the spring axis.
The remaining unit vector orthogonal to $\uvec w$ and $\uvec v$
now follows as
\begin{equation} \label{eq:u}
\uvec u(s) = \uvec v(s) \times \uvec w(s) \\
           = \chi \left( \sin\frac{2\pi s}{\ell} \; \uvx
                       - \cos\frac{2\pi s}{\ell} \; \uvy \right)
           + \sqrt{1-\chi^2} \; \uvz .
\end{equation}
The position of the points located on the perimeter
of the wire's cross-section centered at $\vec r(s)$,
i.e.\ points located on the wire's surface,
can now be expressed with the vector
\begin{equation} \label{eq:s.vector}
\vec c(s,\psi) = \vec r(s) + \frac{d}{2}
\big( \cos\psi \; \uvec u(s) + \sin\psi \; \uvec v(s) \big) ,
\end{equation}
where $\psi$ is the angular coordinate in the wire's cross section,
$0\le\psi<2\pi$, shown in Fig.~\ref{fig:vectors}.
This parameterization is chosen
so that if one fixes the value of $\psi$ and varies $s$,
then one is following a line on the surface of the wire
that runs parallel to the center line of the wire,
i.e.\ that is not winding around the center line
(see the gray dotted curves in Fig.~\ref{fig:vectors}).
For example, if $\psi=\pi/2$ (or $\psi=-\pi/2$)
this curve consists of outermost (or innermost)
elements of the wire relative to the spring's longitudinal axis.
This parameterization of points on the wire surface
is particularly suitable in the present context since,
recalling the assumption that elements on the outer (inner) radius
remain on the outer (inner) radius at all loads,
it follows that the fixed value of $\psi$
always refers to the same elements.
The parameters $s$ and $\psi$ can therefore be understood
as the comoving coordinates on the surface of the spring wire.


\section{Torsion of the spring wire \label{sec:torsion}}

\begin{figure}
\begin{center}
\includegraphics{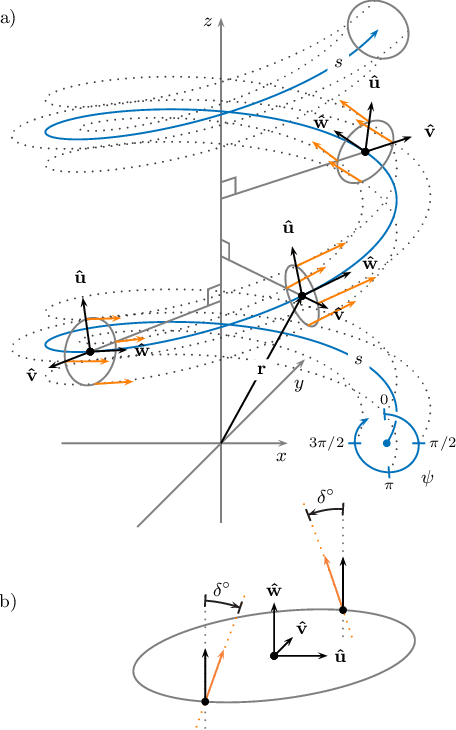}
\end{center}
\caption{\label{fig:vectors} (Color online) Comoving coordinates and vectors:
a) Spring wire cross-sections (gray ellipses)
at three points of the center line.
The spring wire comoving coordinates $s$ and $\psi$ are shown in blue.
The comoving basis vector $\uvec w$
is tangent to the center line,
while $\uvec u$ and $\uvec v$ span the cross-section.
Lines of constant $\psi$
corresponding to topmost ($\psi=0$), outermost ($\psi=\pi/2$), etc.,
elements on the wire's surface are shown with gray dotted curves.
The vectors $\partial \vec c / \partial s$ defined by Eq.~(\ref{eq:dsdsvector})
tangent to these curves are shown in orange.
b) Drawing emphasizes (exaggerated)
that the vectors defined by Eq.~(\ref{eq:dsdsvector})
are not parallel to $\uvec w$.}
\end{figure}

Here we will use Eq.~(\ref{eq:ut.gen})
to compute the contribution of the wire's torsion
to the total elastic potential energy of the spring.
We will consider the curves obtained
by fixing the value of the coordinate $\psi$
in the vector $\vec c(s,\psi)$ defined in Eq.~(\ref{eq:s.vector}).
The angle $\delta^\circ$ that these curves
subtend with the normal to the wire cross-section,
i.e.\ with $\uvec w(s)$ shown in Fig.~\ref{fig:vectors},
is the wire surface's shear angle $\delta^\circ_\shear$
(the superscript $^\circ$ signifies wire surface,
see also Appendix \ref{app:torsion}).
This angle only vanishes in the $\chi=0$ configuration.
To obtain the angle $\delta^\circ$,
we first construct the vectors $\partial \vec c / \partial s$
that are tangential to the curves of constant $\psi$.
After straightforward (but somewhat lengthy) algebraic manipulations,
one can write these vectors using the comoving basis as
\begin{equation} \label{eq:dsdsvector}
\frac{\partial}{\partial s} \, \vec c(s,\psi)
= \frac{d \pi \chi}{\ell} \left( - \sin\psi \; \uvec u(s)
                                 + \cos\psi \; \uvec v(s) \right)
+ \left( 1 + \frac{d\pi}{\ell} \sqrt{1-\chi^2} \sin\psi \right) \uvec w(s).
\end{equation}
The component of $\partial \vec c / \partial s$ of length $d\pi\chi/\ell$
lying in the wire's cross-section
(i.e.~in the plane spanned by $\uvec u$ and $\uvec w$) confirms that,
for $\chi>0$, the vectors $\partial \vec c / \partial s$
are not parallel to $\uvec w$.
The angle $\delta^\circ$ that $\partial \vec c / \partial s$
subtends with $\uvec w$ then is
\begin{equation}
\tan\delta^\circ = \frac{ d \pi \chi / \ell }
                  { 1 + (d\pi/\ell) \sqrt{1-\chi^2} \, \sin\psi },
\end{equation}
which, after expanding in powers of $d/\ell$ and keeping only the leading term,
and for the small angle $\delta^\circ$, gives
\begin{equation}
\delta_\shear^\circ = \frac{d\pi\chi}{\ell} = \frac{d\pi N \chi}{L}.
\end{equation}
Finally, the surface shear angle $\delta_\shear^{\circ}$
of the wire (see Fig.~\ref{fig:torsion}) implies wire torsion
\begin{equation}
\gamma_\tors = \frac{2L}{d} \delta_\shear^{\circ} = 2\pi N \chi .
\end{equation}
The torsion computed above vanishes for $\chi=0$
which means that it represents the torsion of the wire
relative to the inaccessible $\chi=0$ state.
To express the torsional potential energy,
we need the wire torsion relative to the relaxed state
with vanishing shear stress in the wire material,
which is then
\begin{equation}
\gamma_\tors = 2\pi (N\chi - N_0 \chi_0).
\end{equation}
Substituting the above torsion into the expression (\ref{eq:ut.gen})
for the potential energy due to torsion we get
\begin{equation} \label{eq:utorsion}
U_\tors(\chi,N) = G \frac{(d/2)^4 \pi^3}{L}
                     \left( N \chi - N_0 \chi_0 \right)^2 .
\end{equation}
We note that for $N=N_0$ the above expression
is in accord with the spring torsional energy
given by Eq.~(43.7), p.~311, of Sommerfeld.\cite{sommerfeldbook}%
\footnote{In order to connect with Sommerfeld's expression
the following notational changes need to be made:
$M_T \to F(x) D/2$, $\ell\to L$, $\mu\to G$,
and $J_p\to (d/2)^4\pi/2$.  Also $W_T = F(x)^2/2k = k x^2 / 2$ is used.}


\section{Bending of the spring wire \label{sec:bending}}

In order to express the spring wire's bending potential energy,
$U_\bend$ (Eq.~(\ref{eq:ub.gen})),
we first need to express the curvature radius $R$
of the helix representing the center line of the wire
(see Fig.~\ref{fig:bending} and Appendix \ref{app:bending}).
We begin by computing the angle $\Delta\varphi$
subtended by the vectors $\uvec w$ at two points separated by $\Delta s$,
as this is the angle that the wire element
of length $\Delta s$ subtends as seen from its center of curvature.
$\Delta\varphi$ is given by:
\begin{alignat}{1}
\sin\Delta\varphi
& = \left| \uvec w(s) \times \uvec w(s+\Delta s) \right| \notag \\
& = \sqrt{ 2 \left( 1 - \chi^4 + \left( 1 - \chi^2 \right)^2
\cos \frac{2\pi \Delta s}{\ell} \right) } \; \sin \frac{\pi \Delta s}{\ell},
\end{alignat}
which for small $\Delta s$ reduces to
\begin{equation}
\Delta \varphi = \frac{2\pi\Delta s}{\ell} \sqrt{ 1 - \chi^2 }.
\end{equation}
Since the length of the wire element $\Delta s$
can also be expressed as $R\,\Delta\varphi$, we have
\begin{equation} \label{eq:rcurv}
R = \frac{\Delta s}{\Delta \varphi}
  = \frac{\ell}{2 \pi \sqrt{1-\chi^2}}
  = \frac{L}{2 \pi N \sqrt{1-\chi^2}}.
\end{equation}
Using the above result,
the spring wire curvature radius $R_0$
corresponding to the relaxed state of the spring
in which the number of turns and the pitch parameter are $N_0$ and $\chi_0$,
can be written as
\begin{equation} \label{eq:rcurv0}
R_0 = \frac{L}{2\pi N_0 \sqrt{1-\chi_0^2}}.
\end{equation}
Substituting (\ref{eq:rcurv}) and (\ref{eq:rcurv0}) into (\ref{eq:ub.gen})
we obtain
\begin{equation} \label{eq:ubending}
U_\bend(\chi,N) = E \frac{(d/2)^4\pi^3}{2L}
             \left( N \sqrt{1-\chi^2} - N_0 \sqrt{1-\chi_0^2} \right)^2 ,
\end{equation}
representing the potential energy due to bending.
If in the above expression one sets $\chi=\chi_0=0$,
then this agrees with the spring bending energy
given by equation (43.9), p.~311, of Sommerfeld.\cite{sommerfeldbook}%
\footnote{In order to compare with Sommerfeld's expression we use:
$M_B \to \tau(\varphi)$, $\ell\to L$,
$I\to (d/2)^4\pi/4$, and $W_B = \tau^2/2\kappa = \kappa \varphi^2 / 2$.}


\section{Spring potential energy \label{sec:uspring}}

\begin{figure}
\begin{center}
\includegraphics[]{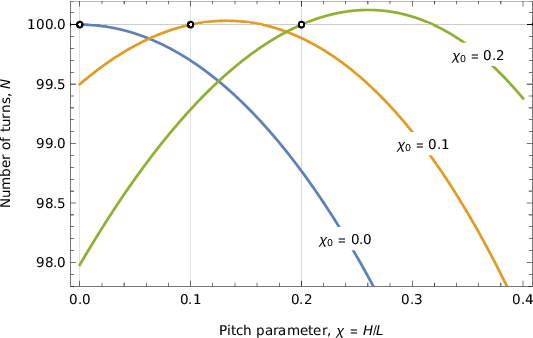}
\end{center} \caption{\label{fig:unwinding}
Unwinding of springs with freely rotating ends:
The number of turns $N$ (Eq.~(\ref{eq:nn.free}))
in springs that have $N_0=100$ turns (indicated with hollow circles)
for $\chi_0 = 0, 0.1, 0.2$.
The ratio of elasticity moduli is taken as $E:G = 5:2$
which is appropriate for many steel alloys.}
\end{figure}

The elastic potential energies due to torsion (Eq.~(\ref{eq:utorsion}))
and due to bending (Eq.~(\ref{eq:ubending})) can now be combined:
\begin{equation} \label{eq:uspring}
U(\chi,N) = \frac{ (d/2)^4 \pi^3 }{ 2L } \left(
    2 G \big( N \chi - N_0 \chi_0 \big)^2
    + E \left( N \sqrt{1-\chi^2} - N_0 \sqrt{1-\chi_0^2} \right)^2
\right) .
\end{equation}
This expression for the spring's elastic potential energy
accounts for the two degrees of freedom of the moving end.

If the moving end of the spring is rotationally fixed,
i.e.\ if $N = N_0$ for all values of $\chi$,
then Eq.~(\ref{eq:uspring}) reduces to
\begin{equation} \label{eq:ufixed}
U_\fixed(\chi) = U(\chi,N_0) = \frac{ (d/2)^4 \pi^3 N_0^2}{ 2L }
\left( 2G \big( \chi - \chi_0 \big)^2
     + E \left( \sqrt{1-\chi^2} - \sqrt{1-\chi_0^2} \right)^2 \right).
\end{equation}
This expression shows that
the contribution due to the torsion of the wire
(term proportional to $G$) is quadratic in the height of the spring,
$H=\chi L$, implying that the wire torsion contributes linearly
to the restoring force of the spring.
The part of $U_\fixed$ due to bending
(term proportional to $E$) is not quadratic in $H$,
implying non-linear contribution to the force.

If the moving end of the spring is allowed to rotate about the spring axis,
then, for a given $\chi$, the number of turns $N$ will change
in a way that minimizes the potential energy of the spring.
The condition $\partial U(\chi,N) / \partial N = 0$
has only one solution,
\begin{equation} \label{eq:nn.free}
N_\free(\chi) = N_0 \frac { 2G\chi_0\chi + E \sqrt{(1-\chi_0^2)(1-\chi^2)}}
{ 2G\chi^2 + E (1-\chi^2)},
\end{equation}
which represents the minimum of $U$ since $U\to\infty$ for $N\to\pm\infty$.
Substitution of the above result into Eq.~(\ref{eq:uspring})
gives the potential energy of the spring with a freely rotating end,
\begin{equation} \label{eq:ufree}
U_\free(\chi) = U(\chi,N_\free(\chi))
= G \frac{ (d/2)^4 \pi^3 N_0^2 }{ L (1 - ( 1 - 2G/E ) \chi^2) }
\left( \chi \sqrt{1-\chi_0^2} - \chi_0 \sqrt{1-\chi^2} \right)^2.
\end{equation}
This result predicts the winding or unwinding of the spring
with a freely rotating end as the spring extends or compresses.
This effect is shown in Fig.~\ref{fig:unwinding}
for 3 different values of $\chi_0$.
We see that if the spring extends starting from its relaxed state
with $\chi=\chi_0$ and $N=N_0$,
it first slightly winds, and then starts to unwind.
We have no simple explanation for this unintuitive behavior.
However, it is a known effect\cite{wahlbook}
and using our model it can be shown
that the spring unwinds back to $N=N_0$
at $\chi \simeq \chi_0 [(4G/E)(1-\chi_0^2) + 2 \chi_0^2 - 1 ]^{-1}$.

Another scenario of potential interest
is when the free end is rotated out of equilibrium by an external torque,
while the height of the spring is held constant.
This situation is approximately realized in the torsional pendulum
where the twisting of the mass provides the torque on the spring.
The potential energy of the spring in this case is
\begin{multline} \label{eq:uu.twist}
U_{\mathrm{twist}}(\phi)
= U \left( \chi, N_\free(\chi) + {\phi}/{2\pi} \right) \\
= \frac{(d/2)^4 \pi (E - (E-2G) \chi^2)}{8L} \, \phi^2
+ \frac{(d/2)^4 \pi^3 E G N_0^2
   ( \chi \sqrt{1-\chi_0^2} - \chi_0 \sqrt{1-\chi^2})}
  { L( E - (E - 2G) \chi^2 ) } ,
\end{multline}
where $\phi$ is the angle of twist relative to the equilibrium configuration
in which the number of turns is given by Eq.~(\ref{eq:nn.free}).
It holds $\Delta \phi = 2\pi \Delta N$.
The above expression will allow us to derive the corresponding torque 
in the following section.


\section{Spring force and torque \label{sec:force}}

The force corresponding to the potential energy (\ref{eq:ufixed})
of the spring with rotationally fixed ends
can be expressed as a function of the pitch parameter $\chi$,
\begin{multline} \label{eq:f.fixed}
F_\fixed(\chi) = - \frac{\dd}{\dd x} U_\fixed(\chi(x))
  = - \frac{\dd\chi}{\dd x} \frac{\dd}{\dd\chi} U_\fixed(\chi) 
  = - \frac{(d/2)^4 \pi^3 N_0^2}{L^2} \left( 2G (\chi-\chi_0)
    - E\chi\left( 1 - \frac{\sqrt{1-\chi_0^2}}{\sqrt{1-\chi^2}} \right)
\right),
\end{multline}
while the force corresponding to the potential energy (\ref{eq:ufree})
of the spring with a freely rotating end can be written as
\begin{equation} \label{eq:f.free}
F_\free(\chi) = - \frac{\dd}{\dd x} U_\free(\chi(x)) = - 2 C_1
\frac{ (1-C_2) \chi_0 \chi + \sqrt{1-\chi_0^2} \sqrt{1-\chi^2} }
     { \left( 1 - C_2 \chi^2 \right)^2 }
\left( \frac{\sqrt{1-\chi_0^2}}{\sqrt{1-\chi^2}} \chi - \chi_0 \right),
\end{equation}
where
\begin{equation} \label{eq:coefs}
C_1 = G \frac{(d/2)^4 \pi^3 N_0^2}{L^2}, \qquad
C_2 = 1 - \frac{2G}{E}.
\end{equation}
It is worth noting that Eq.~(\ref{eq:f.free})
is in accord with the results
obtained by Ivchenko\cite{ivchenko20} and Wahl\cite{wahlbook}
where the derivation was based on force and torque considerations
instead of energy.
As the above expressions for the forces do not allow for easy interpretation,
to obtain some insight into the nonlinearity of the springs
we proceed to compute the power expansions in $x$.
We first consider the spring with a relaxed height $x_0 = \chi_0 L > 0$.
Expanding (\ref{eq:f.fixed}) and (\ref{eq:f.free}) in powers of $x-x_0$
and keeping only the leading non-linear term, we obtain
\begin{equation} \label{eq:ffixed.car}
F_\fixed(x) = - k_\fixed^{(0)} \left( (x-x_0)
  + \frac{ 3 (1+\nu) \chi_0 }{ 2L (1+\nu\chi_0^2) (1-\chi_0^2) } (x-x_0)^2
  + \mathcal{O}\left((x-x_0)^3\right) \right)
\end{equation}
and
\begin{equation} \label{eq:ffree.car}
F_\free(x) = - k_\free^{(0)} \left( (x-x_0)
  + \frac{ 3 (1+3\nu(1-\chi_0^2)) \chi_0 }{
    2L (1+\nu(1-\chi_0^2))^2 (1-\chi_0^2) } (x-x_0)^2
  + \mathcal{O}\left((x-x_0)^3 \right) \right),
\end{equation}
where $\nu$ is Poisson's ratio defined by $\nu = E/2G - 1$.
The quadratic terms in the expansions
represent the nonlinearities of the springs,
i.e.\ the departures from Hooke's law.
The coefficients of elasticity are given by
\begin{equation}
k_\fixed^{(0)} = G \frac{ d^4\pi^3 N_0^2 (1+\nu\chi_0^2) }
                        { 8 L^3 (1-\chi_0^2) }
\qquad \text{and} \qquad
k_\free^{(0)}  = G \frac{ d^4 \pi^3 N_0^2 (1+\nu) }
                        { 8 L^3 (1+\nu(1-\chi_0^2)) (1-\chi_0^2) }.
\end{equation}
The superscript $^{(0)}$ emphasizes that
these coefficients are specific to the relaxed state of the spring.
They are therefore relevant
for the computation of the small oscillations frequency
of a mass attached to a spring close to its relaxed state.
It is also worthwhile noting that
\begin{equation}
\frac{k_\fixed^{(0)}}{k_\free^{(0)}} = 1 +
\frac{ \nu^2 \chi_0^2 ( 1 - \chi_0^2 ) }{ 1 + \nu } \ge 1,
\end{equation}
which means that the spring with rotationally fixed ends
is stiffer than the spring with rotationally free ends.
This is expected since in the case of rotationally free ends
the rotational degree of freedom is being used
to minimize the spring potential energy,
while this does not occur when the ends are fixed.
We also note that the spring constant $k_\fixed$
with $\chi_0=0$ is in accord with Sommerfeld's equation (43.7), p.~311,
(see footnote after our Eq.\ (\ref{eq:utorsion}))
and with other classical texts.%
\cite{sommerfeldbook,timoshenkobook,lovebook,wahlbook}

For springs with $\chi_0\simeq 0$
(i.e.\ closely-coiled ``classroom'' type springs),
the expansions (\ref{eq:ffixed.car}) and (\ref{eq:ffree.car})
are not adequate since, in this case, $k_\fixed^{(0)} \simeq k_\free^{(0)}$,
and the coefficients in the quadratic terms
reflecting the nonlinearity vanish in both expansions.
To reveal the nonlinearity of spring force,
the expansions must then include the cubic terms in $x$:
\begin{equation} \label{eq:ffixed.lab}
F_\fixed(x) = - G \frac{ d^4 \pi^3 N_0^2 }{ 8 L^3 }
                  \left( x + \frac{1+\nu}{2L^2} x^3 + \mathcal{O}(x^5) \right)
\end{equation}
and
\begin{equation} \label{eq:ffree.lab}
F_\free(x) =  - G \frac{ d^4 \pi^3 N_0^2 }{ 8 L^3 } 
                  \left( x + \frac{ 2 \nu }{ L^2 (1 + \nu) } x^3
                           + \mathcal{O}(x^5) \right) .
\end{equation}
The linear coefficients in the two expansions are equal,
while the ratio of the coefficients of the cubic terms is $(1+\nu)^2/4\nu$,
which is greater than one for realistic values of $\nu$.
This again reflects the fact that the spring with rotationally fixed ends
is stiffer than the same spring with freely rotating ends.

For the twisted spring of constant height (Eq.~(\ref{eq:uu.twist})),
the torque required to twist the spring's end out of equilibrium is
\begin{equation} \label{eq:tau.twist}
\tau_{\mathrm{twist}}(\phi)
           = - \frac{\partial}{\partial\phi} U_{\mathrm{twist}} (\phi)
           = - \frac{\pi (d/2)^4 (E - (E - 2G) \chi^2)}{4L} \, \phi ,
\end{equation}
which is linear in the twist angle $\phi$.
The coefficient multiplying $\phi$ in Eq.~(\ref{eq:tau.twist})
is sometimes called the rotational spring constant.
Referring to the rotationally fixed ends situation,
one can also use the general expression
for the potential energy Eq.~(\ref{eq:uspring})
to compute the torque that must be applied to the spring end
to prevent it from rotating,
\begin{multline}
\tau_{\mathrm{fixed}}(\chi)
= \frac{\partial}{\partial\phi} U(\chi,N_0 + \phi/2\pi ) \Big|_{\phi=0} 
= \frac{(d/2)^4 \pi^2 N_0}{2L}
\left( 2G \chi (\chi - \chi_0) + E \left( 1-\chi^2 
  - \sqrt{(1-\chi^2)(1-\chi_0^2)} \right)
\right) .
\end{multline}
Note that there is no negative sign in front of the derivative
because we are computing the torque acting on (not by) the spring.
The derivative is evaluated at $\phi=0$
since we assumed that the moving end does not rotate.


\section{Experimental verification \label{sec:experiment}}

\begin{figure}
\begin{center}
\includegraphics{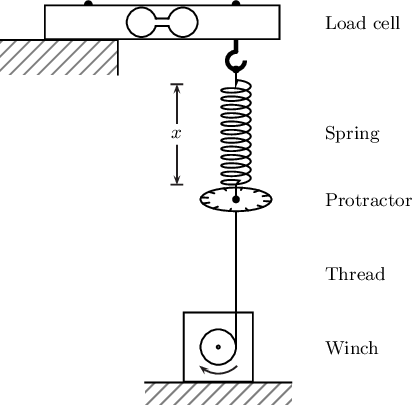}
\end{center}
\caption{\label{fig:exp-setup} The experimental setup:
The load cell is continuously measuring the force
which is read using a computer.
The stepper motor is winching the spring at a constant rate.
The long thin thread between the spring and the winch
allows the lower end of the spring
to rotate freely as the spring extends.}
\end{figure}

\begin{table}
\caption{\label{tbl:springs}
Characteristics of the different springs used in the experiments:
Directly measured parameters and the value of $L$ computed from those
are in the upper part of the table.
The middle part of the table contains the parameters
estimated from the least-squares fit of the models
to the force measurements.
The bottom part of the table contains the values
of the elasticity moduli derived from the model Eq.~(\ref{eq:f.free}).}
\begin{center} \small
\begin{tabular}{ c | c c c c } 

Spring                               & S1                  & S2                   & S3                     & S4 \\

\hline

$d$ [$\mathrm{mm}$]                  & $ 0.40 \pm 0.01 $   & $ 0.60 \pm 0.01 $    & $ 0.80 \pm 0.01 $      & $ 0.90 \pm 0.01 $ \\
$D^{(\mathrm{ext})}_{\mathrm{mcs}}$ [$\mathrm{mm}$]
                                     & $ 8.1 \pm 0.1 $     & $ 13.5 \pm 0.1 $     & $ 14.5 \pm 0.1 $       & $ 15.0 \pm 0.1 $ \\
$H_{\mathrm{mcs}}$ [$\mathrm{mm}$]   & $ 62.0 \pm 0.4 $    & $ 36.2 \pm 0.4 $     & $ 39.0 \pm 0.4 $       & $ 79.35 \pm 0.4 $ \\
$N_{\mathrm{mcs}} \simeq N_0$        & $ 147.0 \pm 0.3 $   & $ 57.5 \pm 0.3 $     & $ 46.5 \pm 0.3 $       & $ 87.0 \pm 0.3 $ \\
$m$ [$\mathrm{g}$]                   & $ 3.7 \pm 0.1 $     & $ 5.7 \pm 0.1 $      & $ 8.5 \pm 0.1 $        & $ 19.7 \pm 0.01 $ \\
$L$ [$\mathrm{m}$]                   & $ 3.56 \pm 0.05 $   & $ 2.33 \pm 0.02 $    & $ 2.00 \pm  0.02 $     & $ 3.86 \pm 0.03 $ \\

\hline

$\bar k$ [$\mathrm{N}/\mathrm{m}$]   & $ 4.058 \pm 0.002 $ & $ 10.97 \pm 0.01  $  & $ 33.70 \pm 0.03 $     & $ 28.79 \pm 0.01 $ \\
$C_1$ [$\mathrm{N}$]                 & $ 7.12  \pm 0.04 $  & $ 12.6 \pm 0.1 $     & $ 33.2 \pm 0.3 $       & $ 55.3  \pm 0.3 $ \\
$C_2$ [$10^{-3}$]                    & $ 209   \pm 3 $     & $ 150 \pm 3 $        & $ 153 \pm 4 $          & $ 210 \pm 20 $ \\
$\chi_0$ [$10^{-3}$]                 & $ 9.2   \pm 0.1 $   & $ 0.4 \pm 0.2      $ & $ 7.5 \pm 0.2 $        & $ 10.6 \pm 0.1 $ \\

\hline

$G$ [$\mathrm{GPa}$]                 & $ 84 \pm 9 $        & $ 82 \pm 6 $         & $ 77 \pm 4 $           & $ 85 \pm 4 $ \\
$E$ [$\mathrm{GPa}$]                 & $ 213 \pm 23 $      & $ 193 \pm 14 $       & $ 183 \pm 10 $         & $ 217 \pm 12 $ \\

\end{tabular}

\end{center}
\end{table}

\begin{figure}
\begin{center}
\includegraphics[angle=-90,width=4.5in]{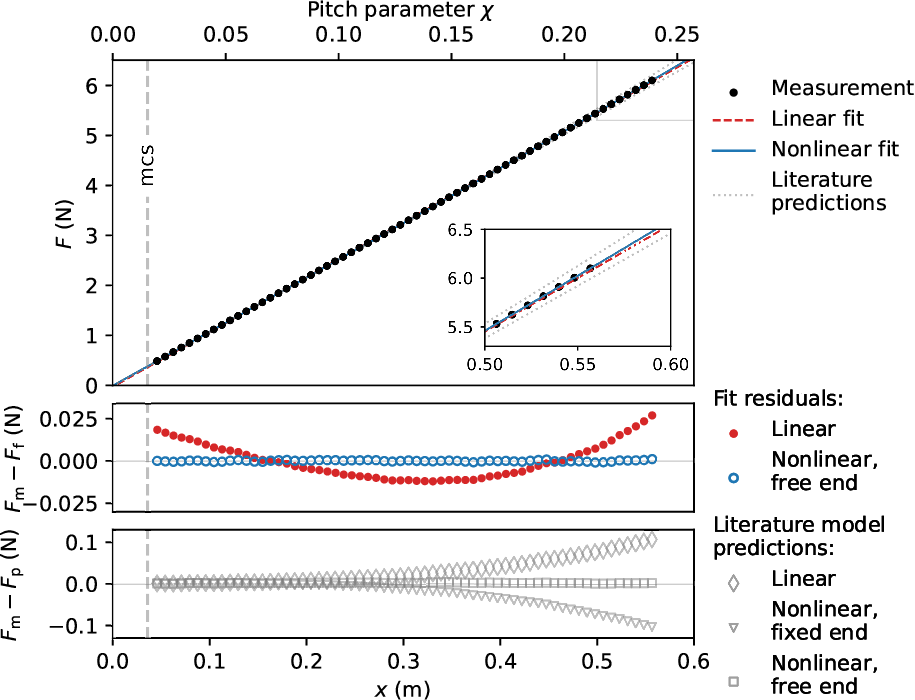}
\end{center}
\caption{\label{fig:exp-ana} (Color online) Top panel:
measured spring force vs.~spring height or pitch parameter
(black dots) for spring S2 (see Table~\ref{tbl:springs}).
The red dashed and blue solid lines are least-squares fits
using the linear and non-linear (free end) models, respectively.
Gray dotted lines are predictions from the literature,
computed using spring dimensions and elastic moduli
determined for S2 in our experiment.
Middle panel: Residuals between the measurements and fits from the top panel. 
Bottom panel: Residuals between the measurements
and literature-based predictions shown in the top panel.
}
\end{figure}

Here we experimentally verify
some of the results presented in the previous sections,
namely the expressions (\ref{eq:nn.free}) and (\ref{eq:f.free}),
giving the unwinding and the force of springs with freely rotating ends.
Fig.~\ref{fig:exp-setup} shows our experimental setup
for directly measuring the force
exerted by the spring at a controlled spring height.
The spring was suspended from the load-cell
that measured the downward force and sent the data to the computer.
The upper end of the spring was fixed both longitudinally and rotationally.
A long thin thread was attached to the lower moving end of the spring.
The thread was then wound around the shaft of the stepper motor
so that the spring could be elongated while allowing its extremity
to rotate freely at all times.
The setup also includes a lightweight protractor
for the measurement of the spring's end rotation,
see Fig.~\ref{fig:protractor}.
Since the motion of the stepper motor can be precisely controlled,
the lower end of the spring can be set to move at the desired constant rate.
All the components that we used in the setup
are standard items available at reasonable cost.
Further technical details regarding the experimental
setup are given in Appendix~\ref{app:setup}

We used springs whose relaxed state
is below their maximally compressed state, i.e.\ $\chi_0 < \chi_\mcs$.
We measured their wire diameter $d$,
the external diameter of the spring body $D_\mcs^{(\mathrm{ext})}$,
the spring height $H_\mcs$,
and the number of turns of the wire $N_\mcs$.
We note that the difference of the number of turns $N_\mcs$
measured in the maximally compressed state
and the number of turns $N_0$ in the relaxed state
   (not experimentally accessible)
is expected to be less than the measurement uncertainty of $N_\mcs$
(see Fig.~\ref{fig:unwinding}),
so that in all calculations we used $N_0 \simeq N_\mcs$.%
\footnote{
   The springs have two states of interest here,
   the maximally compressed state ($\chi_\mcs$, $N_\mcs$),
   and the relaxed state ($0 \le \chi_0 < \chi_\mcs$, $N_0$).
   One can use Eq.~\ref{eq:nn.free} to show that,
   for the given values of $\chi_\mcs$ and $N_\mcs$,
   the difference $\Delta N = N_0 - N_\mcs$ is maximal for $\chi_0=0$,
   and is given by
        $ (\Delta N)_{\max} = N_\mcs
        ( 2 G \chi_\mcs^2 + E ( 1 - \chi_\mcs^2 - \sqrt{ 1 - \chi_\mcs^2 } ) )
        / E\sqrt{1-\chi_\mcs^2} $.
   Taking the spring S1 from Table~\ref{tbl:springs} as an example,
   we have $\chi_\mcs = 0.017$, $N_\mcs = 147$,
   and using $G=80\,\mathrm{GPa}$, $E=200\,\mathrm{GPa}$,
   we find $\Delta N = 0.013$,
   which is by more than one order of magnitude
   smaller than the estimated uncertainty of measurement of $N_\mcs$.
}
The results of the measurements are given in Table~\ref{tbl:springs}.
The length of the wire $L$ is computed by first
subtracting the wire diameter $d$ from the measured
external diameter $D_\mcs^{(\mathrm{ext})}$ of the spring
to obtain the diameter $D_\mcs = D_\mcs^{(\mathrm{ext})} - d$
of the helix representing the wire center line,
and then using the measurements of $H_\mcs$ and $N_\mcs$
to compute $L = N_\mcs \sqrt{ (D_\mcs \pi)^2 + (H_\mcs/N_\mcs)^2 }$.%
\footnote{Although not really needed, the spring mass $m$ was also measured
and was used to test the consistency with other quantities, $d$ and $L$,
through the relation $m = (d/2)^2 \pi L \rho_{\mathrm{steel}}$,
using the standard value of the mass density for steel
$\rho=7850\,\mathrm{kg}\,\mathrm{m}^{-3}$.
For all springs from Table~\ref{tbl:springs}
the agreement between the measured and the predicted mass
was within $10\%$ of the value.}
The spring force vs.\ spring height $x$,
computed from the length of the thread wound on the winch,
is then measured using the setup described above.
A typical measurement for spring S2
is shown in the top panel of Fig.~\ref{fig:exp-ana}.

First, the experimental data could be compared
to the linear regime (Hooke's law).
The best fit is given by the dashed red line
in the top panel of Fig.~\ref{fig:exp-ana},
and its slope is given as the coefficient $\bar k$ in Table~\ref{tbl:springs}.
The position of the zero crossing of the fitted line on the $x$-axis
confirms that the relaxed state of this spring
lies below its maximally compressed state.
In this graph the nonlinearity of the spring is not visible to the naked eye.
However, the residuals of the measurements
with respect to the fitted line reveal a small nonlinearity
shown with filled circles in the middle panel of Fig.~\ref{fig:exp-ana}.

In order to test the consistency of the observed nonlinearity
with the free-end model (\ref{eq:f.free})
we computed the least-squares fits for each of the springs,
treating $C_1$, $C_2$, and $\chi_0$ as free parameters,
and using $\chi$ as the independent variable.
The best fit parameters and their uncertainties
are given in Table \ref{tbl:springs},
where the latter are the combination of
the uncertainties coming from the least-squares fit
and the uncertainties that result form the possible systematic shift
in the independent variable $\chi$,
due to the uncertainties of $L$ and $H_\mcs$ used to compute $\chi$.
Solving (\ref{eq:coefs}) for the elasticity moduli gives
$G = {L^2 C_1}/{d^4 N_0^2 \pi^3}$ and $E = {2G}/{(1-C_2)}$.
These expressions were used to compute the values
of $G$ and $E$ given in Table~\ref{tbl:springs}.
The weighted averages of the elasticity moduli
obtained for the four springs,
$G=(82\pm3)\,\mathrm{GPa}$ and $E=(201\pm15)\,\mathrm{GPa}$,
are consistent with realistic values for spring steel,
$G \sim 80\,\mathrm{GPa}$ and $E \sim 200\,\mathrm{GPa}$.%
\cite{brandesbrook}
The residuals of fit of the model (\ref{eq:f.free}) to the data
are shown with hollow circles in the middle panel of Fig.~\ref{fig:exp-ana}.
They are considerably smaller than those of the linear fit
and they no longer show the trend
which is due to the nonlinearity of the spring.%
\footnote{
Our model does not account for spring's own weight
leading to non-uniform coil pitch when the spring is vertical.
In our experiment, a partial compensation for this effect can be made
by subtracting one half of the spring's weight
from the force measured by the load cell at the top end of the spring.
We have confirmed that with such compensation the results
for $G$ and $E$ given in Table~\ref{tbl:springs} remain unchanged.
}
We also fitted the data with our fixed-end model (\ref{eq:f.fixed})
and found that the residuals are not significantly larger
than with the model (\ref{eq:f.free}),
the value of $G$ remains unaffected,
but the value of $E$ obtained in this fit
is smaller by approximately a factor of $1/2$.

We now wish to compare our energy based derivation of the force
with the models derived via the traditional approaches in the literature.
To compare to the models from the literature,
we computed the spring force using the spring dimensions
and the elastic moduli from Table~\ref{tbl:springs}.
The three classes of models whose residuals are
shown in the bottom panel of Fig.~\ref{fig:exp-ana} are: linear models
from Refs~\onlinecite{lovebook,timoshenkobook,mohazzabi89} (diamonds),
non-linear models with fixed ends from this work
and from Ref.~\onlinecite{wahlbook} (triangles),
and non-linear models with free end from this work
and from Refs~\onlinecite{wahlbook,ivchenko20} (squares).
It turns out that within each class
the models are indistinguishable from each other
so just one model from each class was chosen.
Fig.~\ref{fig:exp-ana} illustrates
that our results are in excellent agreement with the traditional approaches,
the best agreement with the data coming from the the non-linear free-end model.

\begin{figure}
\includegraphics[]{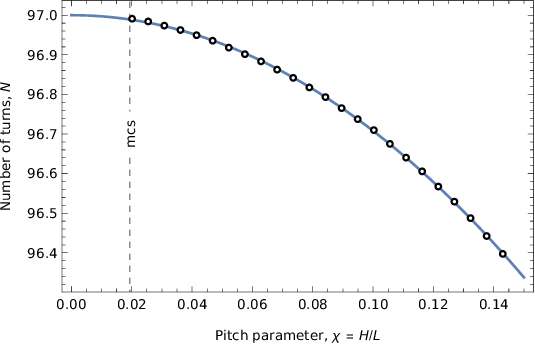}
\caption{\label{fig:exp-unw}
Unwinding of the spring: measurements (dots)
and the model (\ref{eq:nn.free}) (line).
The dashed vertical line indicates the maximally compressed
state of the spring.}
\end{figure}

The experiment was repeated with a spring of geometry S3
in the Table \ref{tbl:springs}, but with $N_\mcs=97$,
and the protractor was used to measure
the angular position of the moving end
with estimated precision of $\pm2.5^\circ$.
The measured unwinding of the spring converted into the number of turns
and the prediction of the model (\ref{eq:nn.free})
are shown in Fig.~\ref{fig:exp-unw}.
The agreement between the data and the model is remarkable,
except for the minor discrepancy that can be seen
at low extensions of the spring.
This discrepancy may be due to the use
of the thin wire approximation in the derivation of the model.


\section{Conclusion \label{sec:conclusion}}

The main result of this paper is the equation (\ref{eq:uspring})
expressing the elastic potential energy of a helical spring
in terms of the spring's two degrees of freedom.
This expression allowed us to compute
the potential energy of the spring with rotationally fixed ends,
as well as with freely rotating ends,
giving us the correction to Hooke's force law in both cases.
The derived expressions were experimentally confirmed
using widely available and affordable equipment,
confirming the correction to Hooke's law.
The model also predicts the winding and unwinding of the spring
with freely rotating ends under varying spring load,
which was experimentally confirmed as well.
The potential energy (\ref{eq:uspring})
can be directly used in the more sophisticated
Lagrangian formalism for the dynamics of a body.
For example, near the end of Sec.~\ref{sec:force}
we already demonstrated how easy it is to apply
Eq.~(\ref{eq:uspring}) to find the torque
exerted by the spring or the rotational spring constant.
Further example could be the motion of the coupled
longitudinal-and-torsional oscillator
that exploits both degrees of freedom of the spring.

We see the full potential of the results presented here
in the design of the physics laboratory setups
demonstrating more advanced phenomena such as transients and resonances
in the oscillatory systems with more than one degree freedom.
In such setups, due to low damping in the mechanical oscillators,
high precision measurements and calculations
are required to sufficiently accurately predict
the frequencies of the normal modes of oscillation.
Some of such experimental setups we may cover in further publications,
where the expressions derived here will help us assess
whether or not the nonlinearity of springs
must be included in the calculations.


\section*{Acknowledgments}

Authors thank the Referees
for suggestions that helped improve the paper.
SI thanks Dr.~Zoran Narančić (UniZG)
for inspiring discussions on elasticity
and for bringing Ref.~\onlinecite{sommerfeldbook} to his attention.


\section*{Author contributions and declarations}

All authors contributed equally to this work
and have no conflicts of interest to disclose.


\bibliography{spring-literature}

\appendix

\section{Elastic energy of a wire under torsion \label{app:torsion}}

If an element of a wire of length $L$ and diameter $d$
is submitted to torsion, see Fig.\ \ref{fig:torsion},
then the small shear angle $\delta^\circ_{\shear}$
at the wire surface (the superscript $^\circ$ signifies surface)
and the wire torsion angle $\gamma_{\tors}$ are related by
\begin{equation} \label{eq:surfshear.gamma}
L \delta^{\circ}_{\shear} = \frac{d}{2} \gamma_{\tors} .
\end{equation}
Assuming that the shear inside the wire
is proportional to the distance $r$ from the wire's center line,
the shear angle inside the wire can be written as
\begin{equation}
\delta_{\shear}(r)
   = \frac{r}{d/2} \delta^{\circ}_{\shear}
   = \frac{r}{L} \gamma_{\tors}, \qquad
   0 \le r \le \frac d 2.
\end{equation}
The elastic potential energy of the wire due to torsion
can now be obtained by integrating the shear potential energy density
\begin{equation}
u_{\shear} = \frac12 G \delta_{\shear}^2,
\end{equation}
over the volume of the wire.
Here, $\delta_{\shear}$ is the shear angle
relative to the relaxed state of the material,
and $G$ is the shear modulus of the material.
Choosing the surface element of the wire cross-section as $\dd S = 2r\pi\,\dd r$
(see the shaded region in the wire cross-section in Fig.\ \ref{fig:torsion}),
we obtain
\begin{multline} \label{eq:u.torsion}
U_{\tors} = \int u_{\shear} \, \dd V
  = \int \frac 12 G \delta_{\shear}^2 (r) \, L \, \dd S 
  = \int_0^{d/2} \frac12 G
    \left( \frac{r\gamma_{\tors}}{L} \right)^2 \, 2r\pi \, \dd r
  = G \frac{(d/2)^4 \pi}{4L} \gamma_{\tors}^2 .
\end{multline}
The above expression assumes
that the wire torsion angle $\gamma_{\tors}$ is measured
relative to the relaxed state of the wire,
where shear stress is zero.
(\ref{eq:u.torsion}) is consistent with the well-known result
for the wire torsional coefficient of elasticity $D=G(d/2)^4\pi/2L$
relating the torque $M$ to torsion $\gamma$ as $M=D\gamma$,
since the potential energy is then
$ U = \int \dd W = \int_0^\gamma M \, \dd \gamma' = D \gamma^2 / 2 $.%
\cite{sommerfeldbook}


\section{Elastic energy of a bent wire \label{app:bending}}

Fig.\ \ref{fig:bending} shows an element of a curved wire
of diameter $d$ and center line length $L_0$.
The curvature radius of the center line is $R$.
Bending the wire entails changing the curvature radius $R$.
The length $L_0$ is assumed to remain constant under changes in $R$,
while the lengths $L_\pm = L_0 \left( 1 \pm (d/2)/R \right)$
of the outermost and the innermost arcs do depend on $R$.
More generally, the arc length at distance $r$ from the center line
can be written as
\begin{equation} \label{eq:arclength}
L(r,R) = L_0 \left( 1 + \frac{r}{R} \right),
\qquad - \frac d 2 \le r \le \frac d 2 .
\end{equation}
If $R_0$ is the curvature radius of the wire center line
in the relaxed state of the wire,
i.e.\ in the state in which the tensile and compressive stress vanish,
and if $R$ is the curvature radius of the bent wire,
then the relative longitudinal deformation
of the wire material can be expressed as
\begin{equation} \label{eq:delta.L.bending}
\delta_L(r) = \frac{L(r,R) - L(r,R_0)}{L(r,R_0)}
            = - \frac{(R-R_0) r}{R R_0 ( 1 + r/R_0 )}
       \simeq - \frac{R-R_0}{R R_0} \, r,
       \qquad - \frac d 2 \le r \le \frac d 2 .
\end{equation}
The approximation made in the last step is appropriate if $d\ll R_0$
and is sometimes referred to
as the ``thin wire approximation.''\cite{ivchenko20}
The elastic potential energy of the wire due to bending
can be obtained by integrating the energy density
\begin{equation}
u_{\tens} = \frac12 E \delta^2_L,
\end{equation}
where $E$ is Young's modulus of the material, over the volume of the wire.
Writing the wire cross-section surface element as
$\dd S = 2\sqrt{ (d/2)^2 - r^2 } \, \dd r$
(see the shaded region in the wire cross-section in Fig.\ \ref{fig:bending})
and also dropping the subscript in $L_0$,
we obtain the potential energy due to bending
\begin{multline} \label{eq:bendint}
U_{\bend}
   = \int u_{\tens} \, \dd V
   = \int \frac12 E \delta_L^2 \, L \, \dd S \\
   = \int_{-d/2}^{d/2} \frac12 E \left( \frac{R-R_0}{R R_0} \, r \right)^2
     L \, 2\sqrt{(d/2)^2 - r^2}  \; \dd r
   = E \frac{(d/2)^4 \pi L}{8 R_0^2} \left( \frac{R-R_0}{R} \right)^2,
\end{multline}
where $R$ is the curvature radius of the bent wire
and $R_0$ is the curvature radius of the wire in its relaxed state.

The simple analysis presented here disregards
the change in $L_0$ that occurs due to bending.
A more elaborate analysis that does account for this phenomenon
can be carried out
by multiplying the arc length (\ref{eq:arclength}) by $1+\epsilon(R)$,
where $\epsilon(R)$ is the relative longitudinal deformation of the center line,
with $\epsilon(R_0) = 0$.
Integration analogous to (\ref{eq:bendint}) is then carried out
and the resulting energy, which now depends on $\epsilon(R)$,
is minimized with respect to $\epsilon(R)$.
The final result can be written as
\begin{equation}
U_{\bend} = E \frac{ (d/2)^4 \pi L R_0^{-2} (R-R_0)^2 }{
  R^2 \left(4 - 2\alpha^2 + 4\sqrt{1-\alpha^2} \right)
- 4 R R_0 \alpha^2 \left( 1 + \sqrt{1-\alpha^2} \right)
+ R_0^2 \alpha^2 \left( 1 + \alpha^2 + \sqrt{1 - \alpha^2} \right) },
\end{equation}
where $\alpha = d / 2R_0$.
In the limit $\alpha \to 0$, which is the thin wire approximation,
the above energy reduces to Eq.~(\ref{eq:bendint}).
This confirms that, within the thin wire approximation,
changes in the wire length can be safely ignored.


\section{Technical details for the experimental setup \label{app:setup}}

\begin{figure}
\begin{center}
\includegraphics[width=3in]{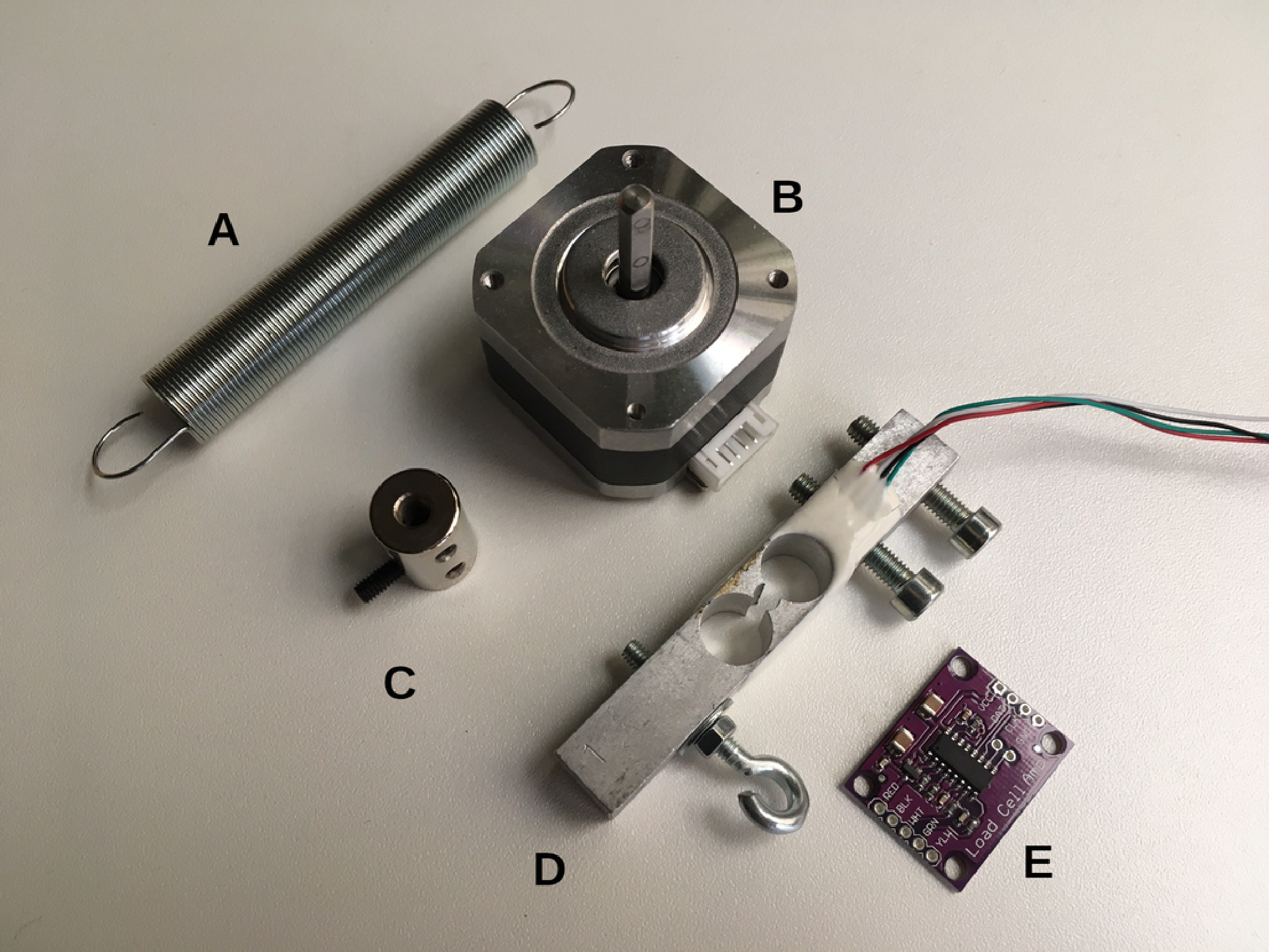}
\end{center}
\caption{\label{fig:hardware} Components used in the experiment:
spring (A), stepper motor (B), shaft coupler (C), load cell (D),
and the break-out board with the load cell signal amplifier
and analog-to-digital converter chip (E).}
\end{figure}

\begin{figure}
\begin{center}
\includegraphics[width=3in]{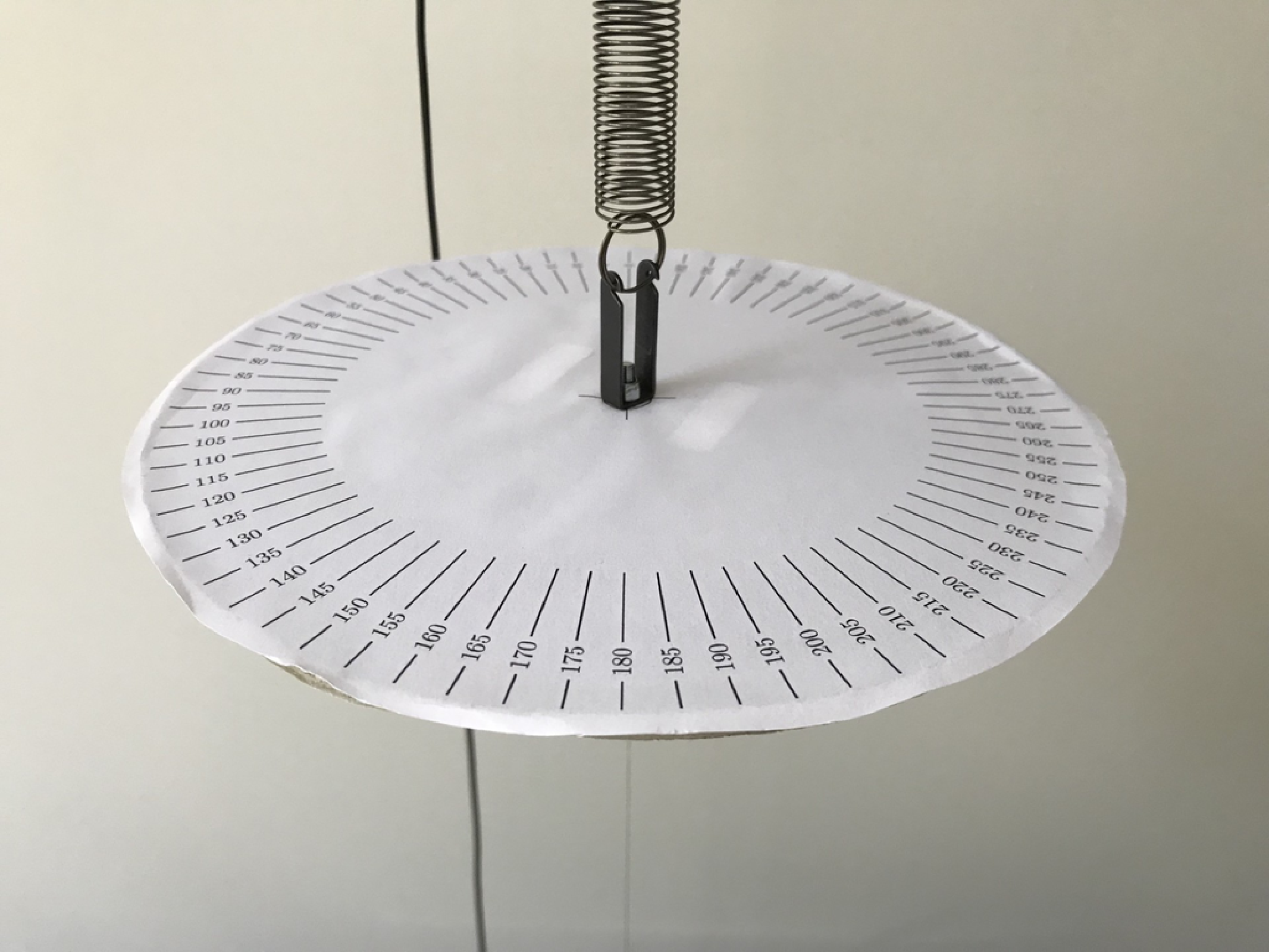}
\end{center}
\caption{\label{fig:protractor}
The protractor printed on paper and glued onto a piece of cardboard:
The protractor is attached to the spring's end so that it rotates with it.
The static wire hanging behind the protractor
(not touching the protractor) helps obtain precise angle readouts.
The thread under the protractor leading to the winch
is barely visible due to its small diameter.
The mass of our protractor is approximately $50~\mathrm{g}$.}
\end{figure}

The experimental setup
is shown schematically in Fig.~\ref{fig:exp-setup},
and some of the components used are shown in Figs~\ref{fig:hardware}
and \ref{fig:protractor}.

We used GALOCE GML611 Micro Load Cell
which is rated for loads up to $1\,\mathrm{kg}$.
It utilizes four strain-gauge resistors
in a Wheatstone bridge configuration.
The load cell is connected to the HX711 load cell signal amplification
and analog-to-digital conversion chip.
For convenience, HX711 can be purchased pre-soldered on a break-out board.
The HX711 chip is further connected to a micro-controller board
from the Arduino family (we used an Arduino Nano)
for which libraries for communication with HX711 are available.

The electric stepper motor that we used for winching
is of standard NEMA~17 size and has 200 intrinsic steps per revolution.
To increase the diameter of the motor shaft we used
the standard shaft coupler of external diameter $14\,\mathrm{mm}$.
The motor is controlled by the TB6600 micro-step driver
making $16$ micro-steps at a frequency of $50\,\mathrm{Hz}$.
Operating the motor at $1\,\mathrm{A}$
provided sufficient torque for the needs of our experiment.
The thread that we used is a braided fishing line
with an approximate diameter of $0.15\,\mathrm{mm}$
and a length of $0.6\,\mathrm{m}$.
The thread material is highly anisotropic;
it is stiff longitudinally, while torsionally it is soft.
The stretch of the thread under a load of $10\,\mathrm{N}$
was found to be smaller than $5\,\mathrm{mm}$,
which is consistent with an estimated $3\,\mathrm{mm}$
using Young's modulus $E=116\,\mathrm{GPa}$
found in the data sheet from Dyneema.\cite{dyneema}
The stretch of the thread during the experiment
could be easily compensated for in the data analysis stage,
but we confirmed that the final results do not change when we do so.
We therefore neglected the stretch of the thread.
We also confirmed that the torque transferred to the spring
via the thread was negligible.
Indeed, we assembled a torsional pendulum using
the softest of our springs and measured the period.
We then repeated the experiment using the thread instead of the spring.
The ratio of the periods was found to be $1:12$,
indicating that the ratio of the torques was $144:1$.

The overall cost of the hardware used for the experiment
(excluding the computer) was below 100\,USD.


\end{document}